\documentclass[a4paper,fleqn,usenatbib]{mnras}

\usepackage[T1]{fontenc}
\usepackage{ae,aecompl}

\usepackage{graphicx}	
\usepackage{amsmath}	
\usepackage{amssymb}	
\usepackage{multirow}

\title[Quasi-periodic oscillations of perturbed tori]{Quasi-periodic oscillations of perturbed tori}

\author[V. Parthasarathy et al.]{Varadarajan Parthasarathy$^{1}$\thanks{E-mail: varada@camk.edu.pl}, Antonios Manousakis\thanks{E-mail: antonism@camk.edu.pl}, W{\l}odzimierz Klu{\'z}niak\thanks{E:mail: wlodek@camk.edu.pl},
\\
$^{1}$N. Copernicus Astronomical Centre, ul. Bartycka 18, 00-716 Warszawa, Poland
}

\date{Accepted 2016 February 06. Received 2016 February 03; in original form 2015 November 2015}

\pubyear{2015}

\begin{document}
\label{firstpage}
\pagerange{\pageref{firstpage}--\pageref{lastpage}}
\maketitle

\begin{abstract}
We performed axisymmetric hydrodynamical simulations of oscillating tori orbiting a non-rotating black hole. The tori in equilibrium were constructed with a constant distribution of angular momentum in a pseudo-Newtonian potential (Klu{\'z}niak-Lee). Motions of the torus were triggered by adding sub-sonic velocity fields: radial, vertical and diagonal to the tori in equilibrium. As the perturbed tori evolved in time, we measured $L_{2}$ norm of density and obtained the power spectrum of $L_{2}$ norm which manifested eigenfrequencies of tori modes. The most prominent modes of oscillation excited in the torus by a quasi-random perturbation are the breathing mode and the radial and vertical epicyclic modes. The radial and the plus modes, as well as the vertical and the breathing modes will have frequencies in an approximate 3:2 ratio if the torus is several Schwarzschild radii away from the innermost stable circular orbit. Results of our simulations may be of interest in the context of high-frequency quasi-periodic oscillations (HF QPOs)  observed in stellar-mass black hole binaries, as well as in supermassive black holes.
\end{abstract}




\section{INTRODUCTION}
\label{section:intro}

Accretion flows in black hole (BH) X-ray binaries and neutron star (NS) low-mass X-ray binaries (LMXBs) have been observed for several decades. Power spectra of the observed light curves have signatures of high-frequency quasi-periodic oscillations (HF QPOs). Stability of the observed frequencies in black hole HF QPOs over long periods of time indicates that they are determined by fundamental features of the system and not accidental quantities such as the flow density or mass accretion rate, as these vary significantly with time. For a black hole accretion system the quantities that do not vary over the  time of observation are the mass and the spin of the central black hole (albeit we do not consider a rotating black hole). Investigating quasi-periodic oscillations (QPOs) is helpful in estimating these two parameters of the central compact body, and in particular the spin \citep{1990ApJ...358..538K, 2001ApJ...548..335S, 2001A&A...374L..19A}. For instance, accretion disk oscillation eigenfrequencies and epicyclic frequencies scale with the mass of the black hole and have a non-trivial dependence on its spin \citep[reviews]{1999PhR...311..259W, 2001PASJ...53....1K, 2013LRR....16....1A}. In many of the accreting black hole binary systems HF QPOs appear in a 3:2 ratio \citep{2001A&A...374L..19A, 2002ApJ...580.1030R}, the mechanism  of which is suggested to be a resonance between orbital and/or epicyclic frequencies of the accretion flow in a strong gravitational field \citep{2001astro.ph..5057K, 2002astro.ph..3314K, 2001AcPPB..32.3605K}. This may also be true in neutron-star LMXBs \citep{2003A&A...404L..21A}. There is no general agreement on the mechanism responsible for quasi-periodic oscillations \citep{2010csxs.book.....L}. 

Oscillatory modes of tori have been investigated analytically and numerically. The first numerical model of oscillating relativistic accretion tori was developed by \citet{2003MNRAS.344L..37R}, HF QPOs in their model  resulting from oscillations of the accretion torus close to a Schwarzschild black hole. \citet{2003MNRAS.344..978R} investigated the gravitational acoustic modes of height-integrated relativistic tori orbiting around a black hole. \citet{2004ApJ...603L..93L} performed numerical simulations of a radially perturbed accretion torus around a compact body  (black hole or neutron star) modelled with the pseudo-potential of \citet{2002MNRAS.335L..29K}, and found vertical epicyclic motion apparently excited by a radial perturbation owing to the non-linear coupling between the epicyclic modes. Light curves and power density spectra (PDS) obtained by ray-tracing photons emitted by a torus oscillating in two eigenmodes were reported by \citet{2004ApJ...617L..45B} who showed that a periodically varying flux of radiation from a luminous torus orbiting a Schwarzschild black hole can result from axisymmetric vertical and horizontal oscillation in the inner regions of accretion flow around black holes. \citet{2006ApJ...637L.113S} demonstrated the possibility of explaining multiple peaks (not observed) in HF QPOs by ray-traced numerical models of an oscillating torus. Several other authors have further numerically investigated QPOs in the Newtonian or in the Paczy{\'n}ski-Wiita pseudo-Newtonian potential \citep{2005MNRAS.357L..31R, 2005MNRAS.362..789R,  2007A&A...467..641S}, and in the relativistic regimes \citep{2005MNRAS.356.1371Z, 2004MNRAS.354.1040M, 2015arXiv151008810M}.

In this study we report on results from Newtonian axisymmetric hydrodynamical simulations of perturbed tori around a Schwarzschild black hole, which is modelled by using the pseudo-potential obtained by \citet{2002MNRAS.335L..29K}, hereafter KL. The motivation for using the KL pseudo-potential stems from the fact that it reproduces the ratio of orbital and epicyclic frequencies for a Schwarzschild black hole.\footnote{This is not the case for the well-known Paczy\'nski-Wiita potential \citep{1980A&A....88...23P}.} The observed peaks of quasi-periodic oscillations could be related to the frequencies of eigenmodes of accretion structures \citep{1999PhR...311..259W, 2001PASJ...53....1K}. Similar velocity perturbations were used in a parallel study based on general relativistic hydrodynamical simulations \citep{2015arXiv151007414M}. The eigenmodes inferred from the simulations of oscillating relativistic tori have correspondence to those identified in the present work. The pseudo-Newtonian simulations reported here are less computationally expensive than the relativistic ones and hence allowed us to clarify certain aspects of the torus response to the assumed perturbations, and in particular to observe in detail the fluid motion in the eigenmodes \citep{2016arXiv160106725P}.

The paper is structured as follows. In Sect.~\ref{section:hydro} we explain the numerical scheme of the code, the initial set-up and describe the models of tori. The results from our simulations are presented in Sect.~\ref{section:results}, discussed in Sect.~\ref{section:disc}, and summarized in Sect.~\ref{section:conc}. 

\section{Hydrodynamic simulations}
\label{section:hydro}

\subsection*{Numerical scheme}

We have used the PLUTO code\footnote{Freely available at http://plutocode.ph.unito.it/} \citep{2007ApJS..170..228M}, a higher order Godunov scheme to integrate system of conservation laws given as
\begin{equation}
\frac{\partial \textbf{U}}{\partial t} + \nabla \cdot \textbf{T}(\textbf{U}) = \textbf{S}(\textbf{U}) ,
\end{equation}
where $\textbf{U}$ is a state vector of conserved variables, $\textbf{T}(\textbf{U})$ is a rank 2 tensor, the rows of which are the fluxes of each component of $\textbf{U}$, and $\textbf{S}(\textbf{U})$ is the source term. We use this scheme to solve the Euler equations of motion in cylindrical coordinates ($r, z, \phi$):
\begin{eqnarray}
&&\frac{\partial \rho}{\partial t} + \nabla \cdot \left(\rho \textbf{v}\right) = 0 \nonumber \\
&&\frac{\partial \rho \textbf{v}}{\partial t} + \nabla \cdot \left(\rho \textbf{vv} + P \textbf{I}\right) = -\rho \nabla \Phi  \nonumber \\          
&&\frac{\partial E}{\partial t} + \nabla \cdot \left[\left(E + P)\textbf{v}\right)\right] = -\rho \textbf{v} \cdot \nabla \Phi	\nonumber
\end{eqnarray}
where $\rho$ is the density, $\textbf{v}$ is the velocity, $P$ is the gas pressure, $\textbf{I}$ is the identity matrix, $E$ the total energy density given as:
\begin{equation}
E = \frac{P}{\gamma - 1} + \frac{1}{2}\rho \left|\textbf{v}\right|^{2}
\end{equation}
where $\gamma$ is the polytropic index and $\Phi$ is the gravitational potential. By exploiting modular structure of the code, spatial order of integration is set to piecewise parabolic method (PPM), dimensionally unsplit second order Runge-Kutta scheme for time evolution and ideal gas equation of state. Roe-type Riemann solver is used for implementing the Godunov scheme. We assumed axisymmetry and uniform grids were used throughout the 2.5D simulations.

\subsection*{Initial profile of torus}

We setup tori around a non-rotating black hole (BH). In order to mimic the effects of general relativity we use the KL potential,

\begin{equation}
\Phi_{\rm KL} = -\frac{GM}{3r_{\rm g}}\left(e^{3r_{\rm g}/R} - 1\right),
\label{eqn:kl}
\end{equation}
where $R = \sqrt{r^{2} + z^{2}}$ is the distance from the center of the compact object, and $r_{\rm g} = 2GM/c^{2}$ is the Schwarzschild radius. The innermost stable circular orbit ($r_{\rm ISCO}$) is at 3$r_{\rm g}$. The KL potential is spherically symmetric and reproduces the ratio of orbital and epicyclic frequencies for a Schwarzschild BH:
\begin{equation}
\omega_{R} = \Omega_{\rm K}\sqrt{1 - \frac{3r_{\rm g}}{R}} ,
\label{eqn:freq}
\end{equation} 
where $\omega_{R}$ is the radial epicyclic frequency and
 $\Omega_{\rm K}\equiv2\pi\nu_{\rm K}$ is the orbital frequency. Note that the vertical epicyclic frequency, $\omega_{\perp}$, is equal to $\Omega_{\rm K}$, as in any spherically symmetric potential. 

The initial distribution of angular momentum in the torus is taken to be
\begin{equation}
\ell = \ell_{\rm c}\left(r/r_{\rm c}\right)^{a}
\label{eqn:ell}
\end{equation}
where $\ell_{\rm c}$ is the orbital angular momentum at $r_{\rm c}$. 
From Eq.~\ref{eqn:kl}
\begin{equation}
\ell_{\rm c} = \sqrt{GMr_{c} e^{3r_{\rm g}/r_{\rm c}}} .
\label{eqn:ellckl}
\end{equation}
The distribution of angular momentum determines the azimuthal velocity, the only non-zero velocity component, which governs the equilibrium configuration of the torus. The profile of surfaces of constant pressure for a torus in hydrostatic equilibrium \citep{1978A&A....63..221A,2000ApJ...528..462H} is given by
\begin{equation}
\Phi_{\rm KL} + \frac{1}{(2 - a)} \frac{\ell^{2}}{r^{2}} + \frac{\gamma}{\gamma - 1}\frac{P}{\rho} = C .
\label{eqn:tor}
\end{equation}
We take $a =$ 0, hence Eq.~\ref{eqn:tor} takes the form
\begin{equation}
\Phi_{\rm KL} + \frac{1}{2} \frac{\ell^{2}_{\rm c}}{r^{2}} + \frac{\gamma}{\gamma - 1}\frac{P}{\rho} = C 
\label{eqn:torprof}
\end{equation} 
where $P =  K \rho^{\gamma}$, $K$ denotes the polytropic constant and we take $\gamma = 5/3$. The constant of integration $C$ is determined from the condition $P =$ 0, that corresponds to zero pressure surface of the torus. The circle of maximum pressure is at $r=r_{\rm c}, z=0$.

\subsection*{Models}

A grid of models is constructed in order to study the behaviour of the torus (see Tab.~\ref{table:property}). The models \verb=T1a-T2b= are \textit{thinner} and the model \verb=T3= is \textit{thicker}. Cross-sectional radii  ($r_{\rm t}$) of the tori are 0.18$r_{\rm g}$ (for \verb=T1a-T2b=) and 0.36$r_{\rm g}$ (for \verb=T3=). All models are scale-free with respect to density and can simulate tori around stellar-mass black holes. A summary of the parameter space is listed in Table.~\ref{table:property}.

\begin{figure*}
\includegraphics[width=2.1\columnwidth]{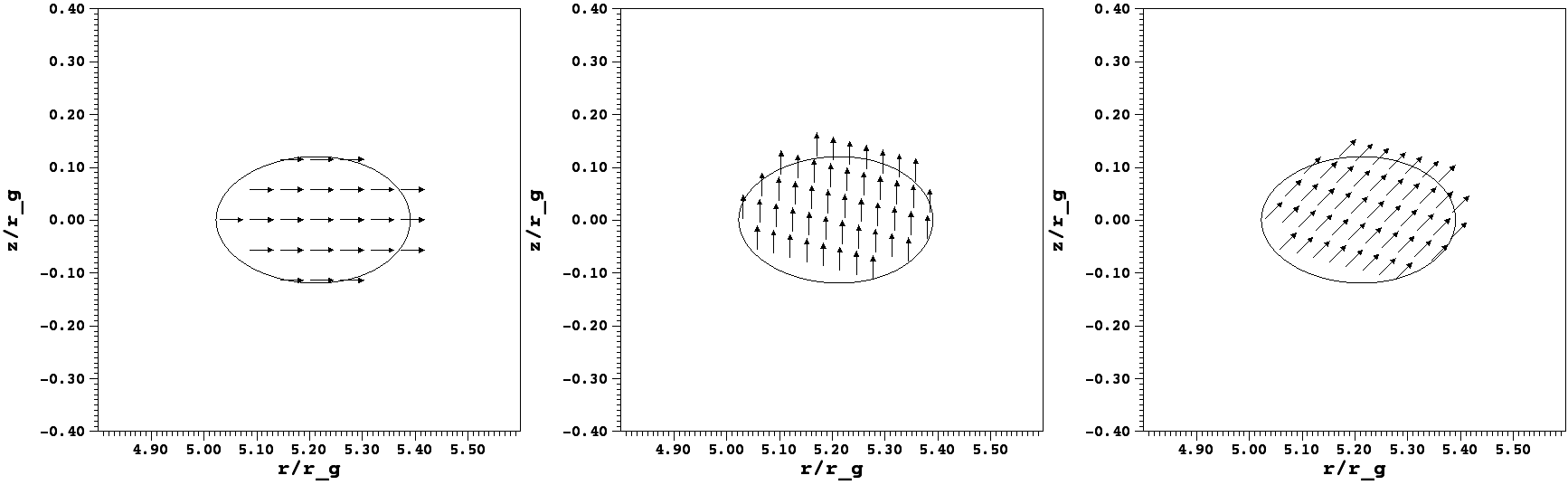}
\caption{Initial configurations of the torus for model T1a. From left to right the velocity perturbations are radial, vertical and diagonal respectively. Magnitudes of the uniform radial and vertical velocity fields are 0.3 in terms of Mach number ($\mathcal{M}$) at $r_{c}$.} 
\label{fig:initial}
\end{figure*}

At $t$ = 0, the tori in equilibrium were perturbed by adding a uniform velocity field. The magnitude of velocity perturbation is always subsonic and constant, 0.3 $\mathcal{M}$ (Mach number), at the center of torus ($r_{\rm c}$). The perturbations are vertical, radial, or diagonal. Initial configurations of the torus for model \verb=T1a= with radial, vertical and diagonal velocity perturbations are shown in Fig.~\ref{fig:initial}.

We perturbed the torus to find its eigenfrequencies.
The  trends of the motion resulting from the velocity perturbations are explained as follows:
\begin{itemize}
\item Vertical perturbation - Initial velocity perturbation is in the $z$ direction. The vertical oscillation occurs at a frequency close to the vertical epicyclic frequency ($\omega_{\perp} = \Omega_{\rm K}$). Oscillations are such that the torus crosses the equatorial plane twice in one orbit. Additional modes are present.
\item Radial perturbation   - Initial velocity is in the $r$ direction. Oscillations are about the initial location of $r_{c}$ on the equatorial plane. The radial oscillation frequency is close to the epicyclic frequency $\omega_{R} < \Omega_{\rm K}$.  Additional modes are present.
\item Diagonal perturbation - Oscillations are both vertical and radial. The torus crosses the equatorial plane twice in one orbit, all the while moving to-and-fro about the initial location of $r_{\rm c}$.
\end{itemize}

\begin{table*}
\caption{The parameter space of the simulations. From left to right the columns are: model name, resolution,  range of the radial ($r$) and vertical ($z$) domain,  inner radius ($r_{\rm in}$),   center of the torus ($r_{\rm c}$),  outer radius ($r_{\rm out}$), cross-sectional radius of the torus ($r_{\rm t}$), magnitude of velocity perturbation (in $\mathcal{M}$ at $r_{\rm c}$) and ratio of the cross-section radius of the torus to the distance from BH ($r_{\rm t}/r_{\rm c}$).  The type  of perturbation is also listed.}
\label{table:property}
\begin{tabular}[width=2\columnwidth]{ccccccccccc}
\hline
\hline
Model  &  Resolution  &  $r/r_{\rm g}$  &  $z/r_{g}$  &  $r_{\rm in}/r_{g}$  &  $r_{\rm c}/r_{\rm g}$  &  $r_{\rm out}/r_{\rm g}$  & $r_{\rm t}/r_{\rm g}$ & $\mathcal{M}$ & $r_{\rm t}/r_{\rm c}$  & Perturbation \\
\hline
\verb=T1a=      &  $296\times228$  &  [4.8 - 5.6]  &  [-0.4 - 0.4]  &  5.02  &  5.2  &  5.39  & 0.18 & 0.3  & 0.03  & vertical, radial, diagonal    \\
\verb=T1b=     &  $512\times496$  &  [4.8 - 5.6]  &  [-0.4 - 0.4]  &  5.02  &  5.2  &  5.39  & 0.18 & 0.3  & 0.03  & vertical, diagonal    \\
\verb=T2a=      &  $296\times228$  &  [6.8 - 7.6]  &  [-0.4 - 0.4]  &  7.02  &  7.2  &  7.39  & 0.18 & 0.3  & 0.02  & vertical, radial, diagonal    \\
\verb=T2b=     &  $512\times496$  &  [6.8 - 7.6]  &  [-0.4 - 0.4]  &  7.02  &  7.2  &  7.39  & 0.18 & 0.3  & 0.02  & vertical, diagonal     \\
\verb=T3=       &  $512\times512$  &  [4.5 - 6.0]  &  [-0.5 - 0.5]  &  4.84  &  5.2  &  5.62  & 0.36 & 0.3  & 0.07  & vertical     \\
\hline
\hline
\end{tabular}
\end{table*}

\section{RESULTS}
\label{section:results}

\subsection{Analysis}

As the perturbed torus evolves in time we measure the $L_{2}$ norm of density defined as
\begin{equation}
\left\Vert \rho \right\Vert_{2} = \left(\sum_{i,j} |\rho_{ij}|^{2}\right)^{1/2}.
\label{eqn:l2n}
\end{equation}
The $L_{2}$ norm ($\left\Vert \rho \right\Vert_{2}$) measures the fluctuations in density of the oscillating torus, in our case corresponding to a combination of eigenmodes triggered by the velocity perturbation.  
The power density spectra of $\left\Vert \rho \right\Vert_{2}$ have been obtained using the Lomb-Scargle technique \citep{1989ApJ...338..277P,1976Ap&SS..39..447L,1982ApJ...263..835S}.

The modes expected from theory for slender tori \citep{2005AN....326..820K, 2006CQGra..23.1689A, 2006MNRAS.369.1235B, 2014A&A...563A.109V} and corresponding values of frequencies from the simulations are given in Table~\ref{table:modes} and Table~\ref{table:modosc} respectively. See Figure 1 of \citet{2006MNRAS.369.1235B} for velocity fields of different modes in the slender torus limit. 

Given the small ratio of $r_{\rm t}/r_{\rm c}$ we expect the model tori to approximately correspond to the slender torus limit (vanishing $r_{\rm t}/r_{\rm c}$). The frequencies of various modes obtained from the simulations agree with those of slender tori (Table~\ref{table:modosc}) with accuracy to the order of $0.03\,\nu_{\rm K}$. The additional frequencies in the simulations other than the fundamental eigenfrequencies are either harmonics mentioned in Table~\ref{table:modosc} (dashed lines in the PDS figures) or discussed below. 

\begin{table}
\caption{Modes and frequencies. From left to right the columns are: modes with labels in parenthesis and theoretical (slender tori) values of eigenfrequencies in units of $\nu_{\rm K}$ at $r_{\rm c}$.}
\label{table:modes}
\begin{tabular}{c|c|c}
\hline
\hline
Modes  & \multicolumn{2}{|c|}{Frequencies}     \\
\hline					
       & $r_{\rm c}$ = 5.2 $r_{\rm g}$ & $r_{\rm c}$ = 7.2 $r_{\rm g}$       \\
\hline
Radial (R)/harmonic  &  0.65/1.3  & 0.76/1.52  \\
Vertical (V)/harmonic ($V_{\rm h}$) & 1.0/2.0  &  1.0/2.0 \\
Breathing (B) & 1.65  & 1.69  \\
Plus (+)      & 0.98  & 1.18  \\
x             & 1.18  & 1.26  \\
\hline
\hline
\end{tabular}
\end{table}

\subsection{Simulations of thinner tori}

\begin{figure*}
\includegraphics[width=18.0cm,height=8.0cm]{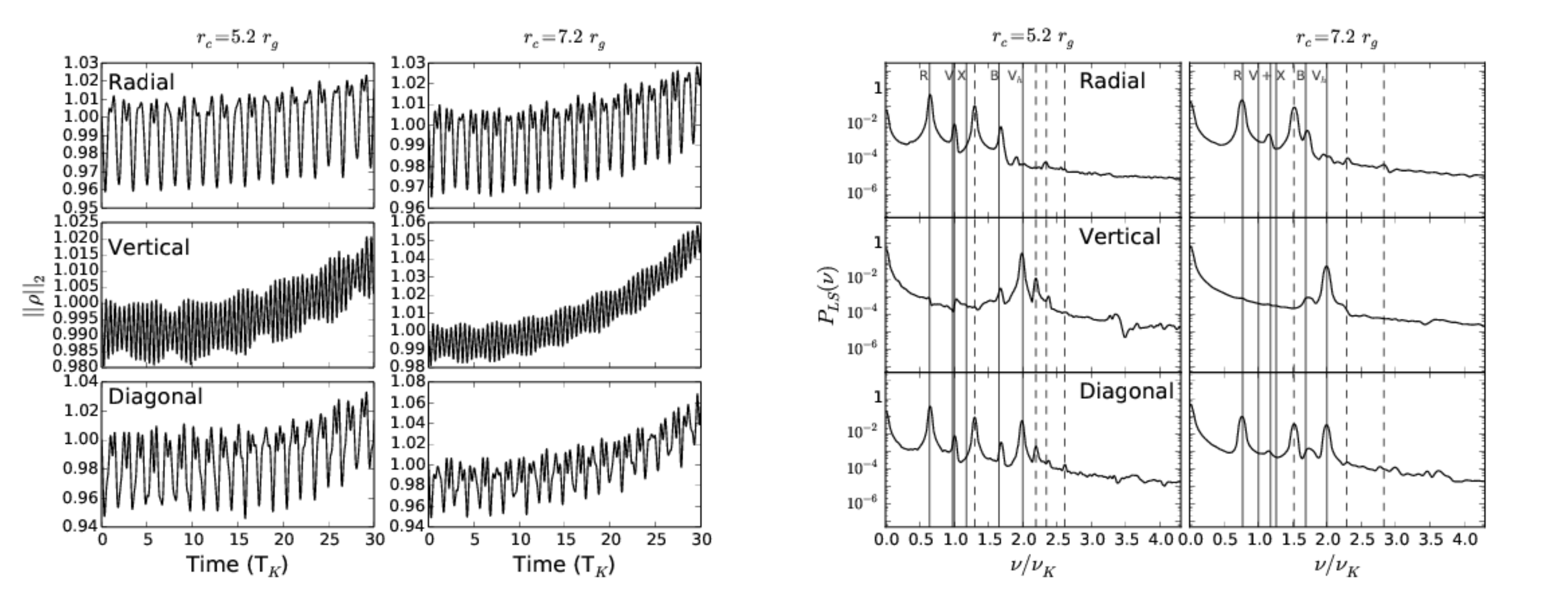}
\caption{\emph{Left:} $\left\Vert \rho \right\Vert_{2}$ as a function of Keplerian time (in units of $T_{\rm K}=1/\nu_{\rm K}$ at $r_{\rm c}$)  for models T1a and T2a. \emph{Right:} The corresponding  PDS of $\left\Vert \rho \right\Vert_{2}$. Frequencies are in units of Keplerian frequency ($\nu_{\rm K}$ at $r_{\rm c}$). The solid lines correspond to the theoretical values of fundamental eigenfrequencies of slender tori modes and dashed lines correspond to harmonics or additional frequencies apparently present in the simulations.}
\label{fig:l2_psd_std}
\end{figure*}

\begin{figure*}
\includegraphics[width=18.0cm,height=8.0cm]{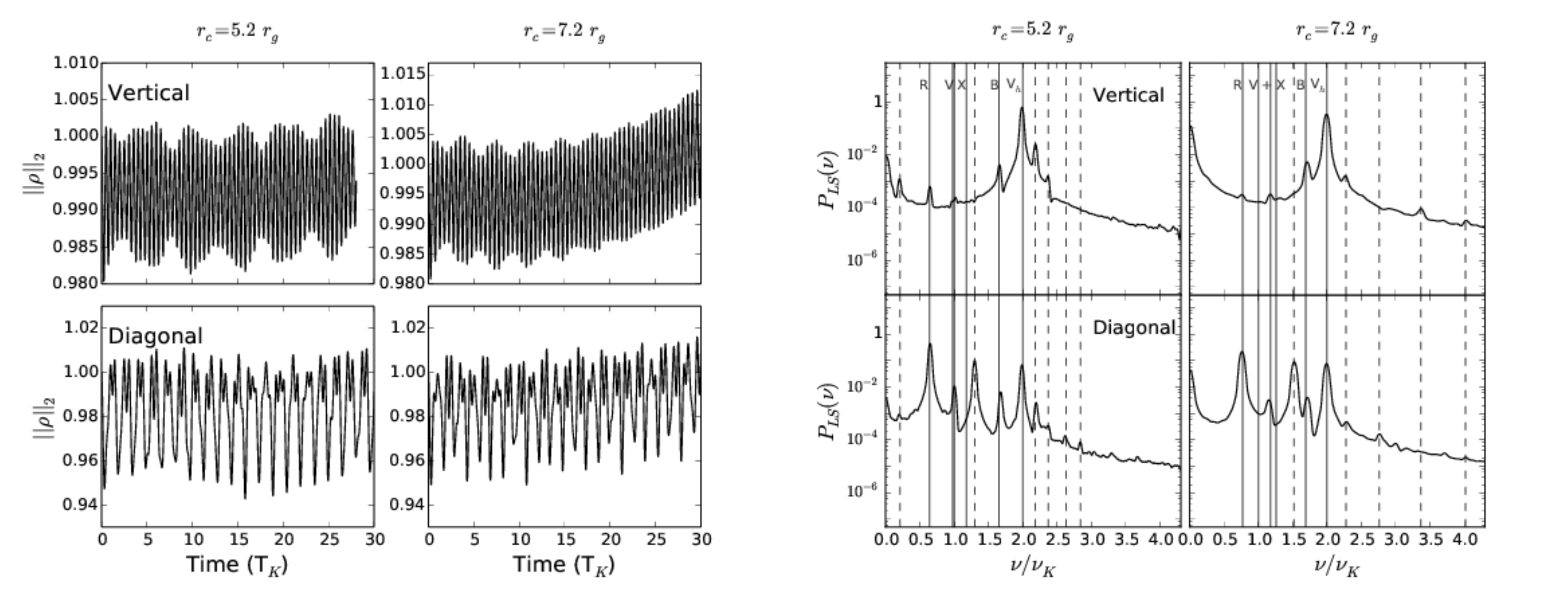}
\caption{Same as Fig. \ref{fig:l2_psd_std} for models T1b and T2b, respectively.}
\label{fig:l2_psd_hr}
\end{figure*}

The motions of models \verb=T1a= and \verb=T2a=  have been triggered with radial, vertical and diagonal perturbations.
Those of models \verb=T1b= and \verb=T2b=  have been triggered with vertical and diagonal perturbations. 
The $\left\Vert \rho \right\Vert_{2}$ and their PDS are shown in Fig.~\ref{fig:l2_psd_std} and Fig.~\ref{fig:l2_psd_hr}, respectively. 

The modes that are excited in the simulations with the radial, vertical and diagonal perturbations (\verb=T1a= and \verb=T2a=) are given in Table~\ref{table:modosc}. 
The radial perturbations excite the radial mode (R) and its harmonic, the plus mode (+) and the breathing mode (B). The frequencies detected in tori excited by the vertical perturbations are the breathing mode (B) and the harmonic of the vertical $V_{\rm h}$. 
It is obvious that diagonal perturbations of the torus excites all the modes of radial and vertical perturbations. Increasing the resolution of computational domain but keeping the other parameters fixed we obtain the models labelled with the letter ``b'' (Tab.~\ref{table:property}).

The modes that are excited in the simulations with  vertical and diagonal perturbations (\verb=T1b= and \verb=T2b=) are given in Table~\ref{table:modosc}. 
Modes excited by the vertical motion are: B and $V_{\rm h}$. Diagonal perturbations excite the modes: R/harmonic, +, B and $V_{\rm h}$.
Minor peaks are at 2.19 and 2.18 for \verb=T1a= and \verb=T1b= respectively, and also at  2.26 for \verb=T2a= and \verb=T2b=. These minor peaks suggests the presence of the x-mode. 

The presence of the radial mode and the plus mode was expected for radial and diagonal oscillation of torus. However, we also observe these modes with weak power for the case of vertical oscillation. We believe that their presence in the latter case is of numerical origin (see Section~\ref{section:unpert}).

\subsection{Simulation of a thicker torus}
\label{section:thick}
\begin{figure*}
\includegraphics[width=2.05\columnwidth]{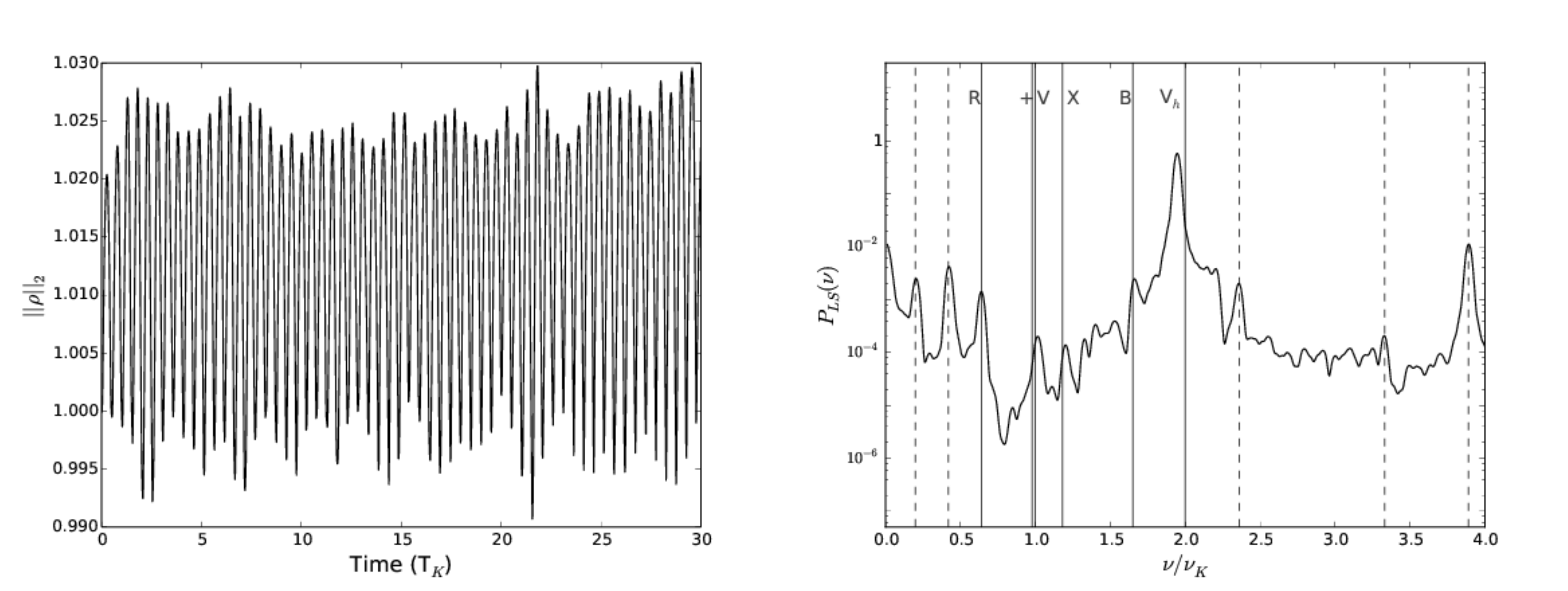}
\caption{Same as Fig. \ref{fig:l2_psd_std} for model T3. Note the strong harmonics of the X-mode (at $2.36\,\nu_{\rm K}$) and of the vertical epicyclic mode at 1.95 and  $3.89\,\nu_{\rm K}$.
}
\label{fig:bigtor}
\end{figure*}

To further investigate, a thicker torus is employed (see model \verb=T3=). However, the initial magnitude of velocity perturbation was kept fixed. 
Simulation for \verb=T3= was performed by perturbing the torus vertically. The resulting $\left\Vert \rho \right\Vert_{2}$ and PDS are displayed in Fig.~\ref{fig:bigtor}. 

All the frequencies and the corresponding modes derived from \verb=T3= are given in Tab.~\ref{table:modosc}. The modes apparently excited by a uniform vertical perturbation of a slightly non-slender torus are: R, +, x/harmonic, B/harmonic and $V_{\rm h}$.

\subsection{Simulations of unperturbed tori}
\label{section:unpert}
\begin{figure*}
\includegraphics[width=18.0cm,height=8.0cm]{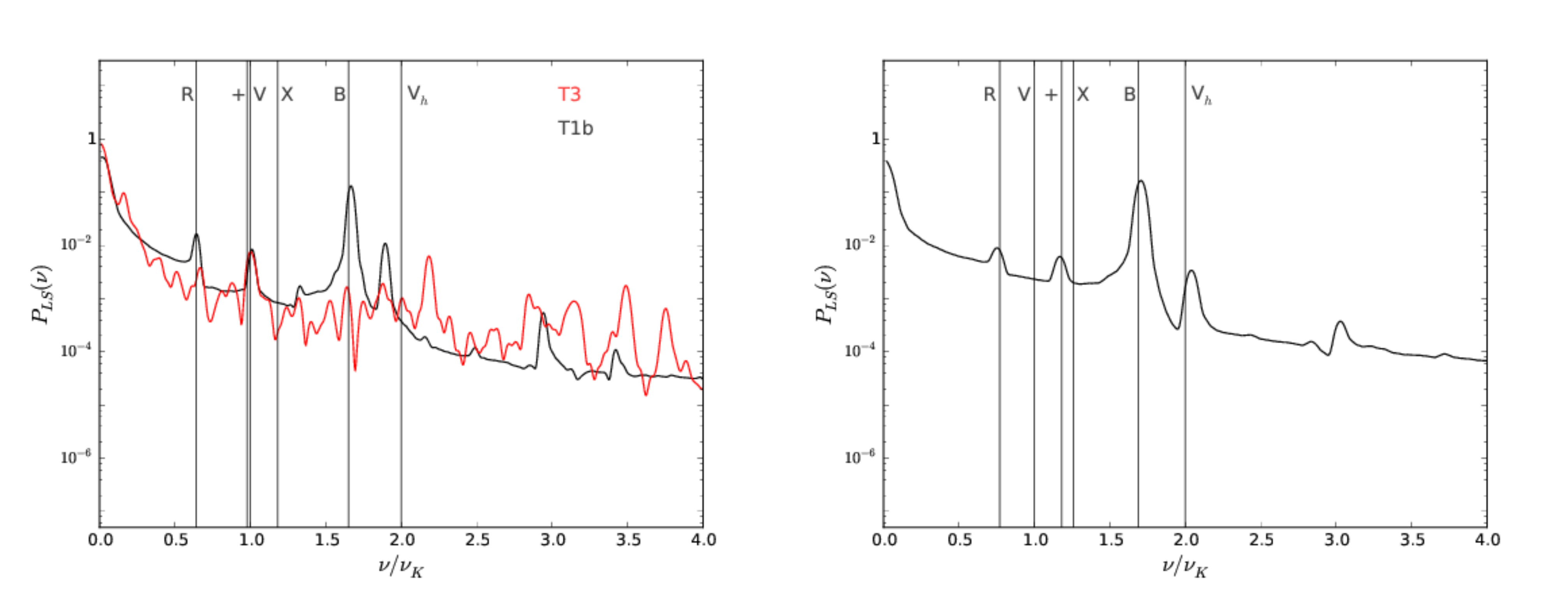}
\caption{\emph{Left:} PDS of $\left\Vert \rho \right\Vert_{2}$ for unperturbed models T1b (black solid line) and T3 (red solid line) respectively. Frequencies are in units of Keplerian frequency, $\nu_{\rm K}$, at $r_{\rm c}$ = 5.2$r_{\rm g}$. \emph{Right:} PDS of $\left\Vert \rho \right\Vert_{2}$ for unperturbed model T2b. Frequencies are in units of Keplerian frequency, $\nu_{\rm K}$, at $r_{\rm c}$ = 7.2$r_{\rm g}$. The solid lines correspond to the theoretical values of fundamental eigenfrequencies in the slender torus limit. The excitation of the various modes in the unperturbed torus is of numerical origin.}
\label{fig:unpert}
\end{figure*}

The radial mode and the plus mode excited from the vertical oscillations of models \verb=T1b=, \verb=T2b= and \verb=T3=, were further investigated by simulating these models without initial velocity perturbations. The PDS of $\left\Vert \rho \right\Vert_{2}$ for unperturbed models \verb=T1b= (black solid line) and \verb=T3= (red solid line) are displayed in the left panel of Fig.~\ref{fig:unpert}. The right panel of Fig.~\ref{fig:unpert} shows the PDS of $\left\Vert \rho \right\Vert_{2}$ for unperturbed model \verb=T2b=. Modes inferred from the simulations of unperturbed tori \verb=T1b=, \verb=T2b= and \verb=T3= are: the radial mode (R), the plus mode (+) and the breathing mode (B). These modes are also present in the vertically perturbed torus. However, we note that the vertical mode is absent in both the unperturbed torus simulation and in the one with the radial perturbation. These result strongly suggest that in the vertically perturbed torus the R, +, and B modes are excited by an interaction of the background gas with the torus, and hence that their presence is of numerical origin.

\begin{table*}
\caption{Frequencies and corresponding modes of oscillating tori identified in simulations. From left to right: model name, type of perturbation, frequencies in units of $\nu_{\rm K}$ at $r_{\rm c}$ obtained from simulations and modes corresponding to solid lines in PDS figures,  respectively. }
\label{table:modosc}
\begin{tabular}[width=2\columnwidth]{cccc}
\hline
\hline
Model  & Trend of perturbation  & Frequencies & Modes  \\
\hline
\verb=T1a=    &  radial   &   0.65/1.3, 1.01, 1.68       &  R/harmonic, +, B     \\
              &  vertical &   1.67, 1.99                 &  B, $V_{\rm h}$                 \\
              &  diagonal &   0.65/1.3, 1.0, 1.68, 1.99  &  R/harmonic, +, B, $V_{\rm h}$  \\
\hline
\verb=T1b=    &  vertical &  0.65, 1.01, 1.66, 1.99       & R, +, B, $V_{\rm h}$          \\
	          &  diagonal &  0.65/1.29, 1.0, 1.68, 1.99   & R/harmonic, +, B, $V_{\rm h}$ \\
\hline
\verb=T2a=    &  radial   &   0.77/1.53, 1.15, 1.71       & R/harmonic, +, B    \\
	          &  vertical &   1.71, 1.99                  & B, $V_{\rm h}$                \\
	          &  diagonal &   0.77/1.52, 1.16, 1.73, 1.99 & R/harmonic, +, B, $V_{\rm h}$ \\
\hline
\verb=T2b=    &  vertical &  0.76, 1.18, 1.71/3.37, 1.99/4.03  & R, +, B/harmonic, $V_{\rm h}$/harmonic  \\
              &  diagonal &  0.76/1.52, 1.16, 1.72, 2.0        & R/harmonic, +, B, $V_{\rm h}$   \\
\hline
\verb=T3=     &  vertical &  0.64, 1.02, 1.2/2.36, 1.66/3.33, 1.95/3.89 & R, +, x/harmonic, B/harmonic, $V_{\rm h}$/harmonic  \\
\hline
\hline
\end{tabular}         
\end{table*}

\subsection{Summary of results}
\label{section:summary}
The radial mode (R), plus mode (+) and breathing mode (B) were excited by radial perturbations of \textit{thinner} tori \verb=T1a= and \verb=T2a=. The plus mode and radial mode occur in an approximate 3:2 ratio. No vertical motions were excited by a radial perturbation.

The vertical mode ($V_{\rm h}$, see Sec.~\ref{section:disc}) and breathing mode (B) were excited by vertical oscillations of \textit{thinner} tori \verb=T1a= and \verb=T2a=. The radial mode (R), plus mode (+), breathing mode (B) and vertical mode ($V_{\rm h}$) were excited by vertical oscillations of \textit{thinner} tori \verb=T1b= and \verb=T2b=.

The radial mode (R), plus mode (+), x-mode, breathing mode (B) and vertical mode ($V_{\rm h}$) were excited by the vertical oscillation of \textit{thicker} torus \verb=T3=.

The modes inferred from the simulations of unperturbed tori \verb=T1b=, \verb=T2b= and \verb=T3= are: R, + and B. 
Hence, R, + and B modes present in the vertical oscillations are numerical artefacts. There is no coupling between radial and vertical epicyclic motion. 

\section{Discussion}
\label{section:disc}

We have performed axisymmetric hydrodynamical simulations of oscillating tori in the Klu{\'z}niak-Lee potential (Eq.~\ref{eqn:kl}), which reproduces the ratio of orbital and epicyclic frequencies for a Schwarzschild black hole (Eq.~\ref{eqn:freq}). In order to investigate high-frequency quasi-periodic oscillations in accretion disks around black holes, equilibrium tori were constructed with a constant distribution of angular momentum in KL potential and perturbed with uniform sub-sonic velocity fields (radial, vertical and diagonal respectively).  The perturbed tori were allowed to evolve in time and the results obtained are discussed below.

In the limit of a slender torus ($r_{\rm t}/r_{\rm c}\rightarrow 0$) the vertical eigenmode corresponds to an incompressible, rigid motion at the vertical epicyclic frequency, while the radial mode corresponds to a rigid radial oscillation of each cross-section \citep{2004ApJ...617L..45B, 2005AN....326..820K}. It is important to note that our tori have a fairly large cross-sectional radius in the sense of a significant difference in the values of the epicyclic frequencies between the inner and the outer edge of the torus. Hence, the simulated tori have more complicated eigenmodes, including a see-saw motion of the cross-section in the vertical mode \citep[see Fig. 2 of][]{2016arXiv160106725P}, and a radial stretching/compression in the radial mode, corresponding to the theoretical eigenfunctions of a non-slender torus  \citep[see Fig. 5 of][]{2007ApJ...665..642B}. For the same reason, the assumed initial velocity perturbations do not correspond to the actual eigenmodes of non-slender tori. The uniform radial velocity perturbations supplied to the numerical set-up initially can be decomposed into a combination of eigenmodes of  tori, the strongest components being the radial mode and the plus mode. Similarly, a uniform vertical perturbation corresponds to a combination of (mostly) vertical and X modes.

The presence of the breathing mode is inferred from the corresponding frequency peak in the radial, vertical and diagonal oscillations of tori (\verb=T1a= - \verb=T3=) (Figs.~\ref{fig:l2_psd_std}, ~\ref{fig:l2_psd_hr} and ~\ref{fig:bigtor}). However, the conclusions regarding its origin are complicated by the fact that owing to numerical truncation error, and to the interaction of the torus with the background, the equilibrium torus (Eq.~\ref{eqn:torprof}) will not be in perfect equilibrium when discretized on the grid. The torus expands and contracts due to the lack of perfect equilibrium, which likely triggers the breathing mode (B). Indeed, the breathing mode frequency is observed in the simulations of unperturbed tori{ (\verb=T1b=, \verb=T2b= and \verb=T3=) (Fig.~\ref{fig:unpert})}. The breathing mode was observed in oscillating and unperturbed tori in previous Paczy{\'n}ski-Wiita potential \citep{2007A&A...467..641S} and recent general relativistic hydrodynamics numerical simulations \citep{2015arXiv151007414M}. 

The radial oscillations of the torus are about the initial location of the radius of pressure maximum ($r_{\rm c}$) on the equatorial plane. As the uniform initial radial velocity perturbation  corresponds to the theoretical eigenmode of a radially oscillating {\it slender} torus, in the non-slender torus this velocity perturbation contains both the radial and plus mode. Frequencies of the radial epicyclic mode are in accordance with Eq.~\ref{eqn:freq}. The modes excited by radial perturbations of \verb=T1a= and \verb=T2a= (Fig.~\ref{fig:l2_psd_std}) are; R/harmonic, + and B (Tab.~\ref{table:modosc}). The plus mode surface gravity wave \citep{2006MNRAS.369.1235B} was observed in previous numerical simulations \citep{2005MNRAS.357L..31R, 2005MNRAS.362..789R, 2003MNRAS.344L..37R, 2003MNRAS.344..978R, 2005MNRAS.356.1371Z}. The plus mode and the radial mode (Tab.~\ref{table:modosc}) occur in an approximate 3:2 ratio (1.55 and 1.50, respectively) for these locations of the torus \citep{2006MNRAS.369.1235B}. The appearance of the plus mode (+) and the radial epicyclic mode in an approximate 3:2 ratio were proposed by \citet{2003MNRAS.344L..37R} and obtained in other simulations \citep{2007A&A...467..641S, 2015arXiv151007414M}. Vertical epicyclic motion was not excited by radial oscillations of \verb=T1a= and \verb=T2a=. The presence of a radial mode and the plus mode was inferred in the simulations of unperturbed tori \verb=T1b= and \verb=T2b= (Sec.~\ref{section:unpert}, Fig.~\ref{fig:unpert}). The plus mode was previously observed in the simulations of unperturbed tori \citep{2007A&A...467..641S}. 

The torus crosses the equatorial plane twice in one orbital period during vertical oscillations. Considering the fact that the torus is symmetric with respect to the equatorial plane, the density of the torus fluctuates \textit{twice} per cycle, as it reaches the maximum amplitude (in its motion above and below the equatorial plane) and then as it crosses the equatorial plane. For this reason the $L_{2}$ norm reveals the vertical epicyclic oscillation at double the eigenfrequency. The frequency of the vertical mode (V) is approximately equal to the orbital frequency ($\omega_{\perp} \approx \Omega_{\rm K}$), and the  PDS reveal (Figs.~\ref{fig:l2_psd_std}, Fig.~\ref{fig:l2_psd_hr}) that in the $L_2$ norm the power of vertical motion  is  concentrated in the first harmonic ($V_{\rm h}$) at twice the orbital frequency. 
 
The modes excited by vertical oscillations of \verb=T1a= and \verb=T2a= are: $V_{\rm h}$, and B (see Tab.~\ref{table:modosc}). The radial mode and plus mode are present in vertical oscillations of \verb=T1b= and \verb=T2b= (Fig.~\ref{fig:l2_psd_hr}), along with vertical mode ($V_{\rm h}$) and breathing mode (B) (Tab.~\ref{table:modosc}). However from the simulations of unperturbed tori it is evident that the vertical oscillations of \verb=T1b= and \verb=T2b= do not excite radial epicyclic motion, rather it is likely to be a numerical artefact (Fig.~\ref{fig:unpert}). If there were coupling between the radial and vertical epicyclic motion, vertical motion would have been excited by a radial perturbation; this was not observed. 

Minor peaks (dashed lines in Fig.~\ref{fig:l2_psd_std} and Fig.~\ref{fig:l2_psd_hr}) at 2.19, 2.18, 2.26 are suggestive of the x-mode present in the torus. The x-mode is expected to be present because the initial (uniform) vertical velocity perturbation does not correspond to a single eigenmode of the simulated torus of significant thickness, and corresponds to a combination of the vertical mode and the x-mode. Vertical oscillations of a dust torus would be such that its inner radius would complete one cycle of epicyclic motion more quickly than the outer radius would. In an eigenmode of the hydrodynamic torus all motions must of course occur at the same frequency, however different parts of the torus may have different velocities. In fact, we find that in the vertical eigenmode the torus oscillates in such a manner that the torus swings from one extreme displacement to the other in a see-saw fashion \citep[Fig. 2 of][]{2016arXiv160106725P}. This behaviour is in agreement with the pattern of velocity in the vertical epicyclic eigenmode found by \citet{2007ApJ...665..642B}. 

The modes excited by the vertical perturbation of model \verb=T3= (Fig.~\ref{fig:bigtor}) are: R, +, x/harmonic, B/harmonic and $V_{\rm h}$ (see Tab.~\ref{table:modosc}). The harmonic of the x-mode at frequency $2.36\,\nu_{\rm K}$ has significantly more power than the fundamental of the x-mode at frequency $1.2\,\nu_{\rm K}$, which follows from the same reasoning given above for inferring the harmonic of vertical mode. The x-mode was found to be excited in a recent numerical investigation of relativistic slender tori \citep{2015arXiv151007414M}. The excitation of the breathing mode is probably of numerical origin, as explained above. The radial mode and the plus mode are not triggered by the vertical oscillation of \verb=T3= as inferred from the simulation of unperturbed torus \verb=T3= (left panel in Fig.~\ref{fig:unpert}). Here, the power of the radial and breathing modes excited in the simulation of unperturbed torus \verb=T3= are not stronger than noise, presumably due to the bigger size of the torus (left panel in Fig.~\ref{fig:unpert}). The peak located close to the frequency 2$\nu_{\rm K}$ excited in the simulation of unperturbed torus \verb=T2b= (right panel in Fig.~\ref{fig:unpert}), is likely not to be the vertical mode ($V_{\rm h}$), as we do not infer the vertical mode ($V_{\rm h}$) to be excited from the simulations of unperturbed tori \verb=T1b= and \verb=T3= (left panel Fig.~\ref{fig:unpert}). The peaks not corresponding to the theoretical values of fundamental eigenfrequencies of slender tori modes, although apparently present in the simulations of unperturbed tori are presumably numerical artefacts.

Diagonal oscillations triggers all the modes mentioned and explained above for models \verb=T1a=-\verb=T2b=. The modes excited by diagonal oscillations of \verb=T1a=-\verb=T2b= (Fig.~\ref{fig:l2_psd_std} and Fig.~\ref{fig:l2_psd_hr}) are: R/harmonic, +, B and $V_{\rm h}$ (see Tab.~\ref{table:modosc}). 

It is interesting to observe the close similarity of our results with recent simulations performed by \citep{2015arXiv151007414M}. The authors have used similar trends of velocity perturbations to investigate quasi-periodic oscillations in the general relativistic hydrodynamical regime. The correspondence between the results of that study and the one reported here confirms that conclusions about oscillations of relativistic tori can be drawn from Newtonian simulations in the PLUTO code if an appropriate pseudo-potential \citep{2002MNRAS.335L..29K} is used.

\section{Conclusions}
\label{section:conc}

Axisymmetric hydrodynamical simulations of oscillating tori in the Klu{\'z}niak-Lee (2002) potential were performed using the PLUTO code. The parameter space of the simulations spans a grid of models (Tab.~\ref{table:property}). The models represent tori orbiting a black hole. We identified eigenmodes of the most prominent modes of oscillation. This was done primarily relying on the correspondence of the detected frequencies with eigenfrequencies of slender tori as a function of the equilibrium radial position of the torus. The simulations were of sufficiently high resolution and duration that we were able to see the correspondence of the radial and vertical eigenmodes with the theoretical computations of the velocity eigenfunctions of non-slender tori \citep{2007ApJ...665..642B}.

The most prominent peaks in the PDS of radially perturbed tori correspond to the radial mode (R), the plus mode (+) and the breathing mode (B). The plus mode and radial mode occur in an approximate 3:2 ratio for tori located close to the black hole \citep{2006MNRAS.369.1235B}. No vertical motions were excited by the radial perturbation in our study.

For the first time we study numerically the frequencies of oscillation of a torus with an initial vertical motion. We conclude that both the vertical epicyclic mode and the X-mode may be strongly excited in such tori.

Thanks to their superior resolution and duration, the pseudo-Newtonian numerical simulations reported here clarified which modes are excited in general relativistic simulations of tori around black holes 
\citep[compare][]{2006ApJ...637L.113S,2015arXiv151007414M,2015arXiv151008810M}.

\section*{Acknowledgements}

The authors are grateful to the anonymous referee for the helpful comments. Our research was supported by Polish NCN grant 2013/08/A/ST9/00795.




\bibliographystyle{mnras}
\bibliography{qpo_kl_ref} 


\bsp	
\label{lastpage}
\end{document}